\documentclass[twocolumn,aps,showpacs]{revtex4}
\usepackage{amssymb}
\usepackage{epsf}
\usepackage{amsmath}
\usepackage{graphicx}

\def\lsim{\lower -0.3ex \hbox{$<$} \kern -0.75em \lower 0.7ex \hbox{$\sim$}}
\def\gsim{\lower -0.3ex \hbox{$>$} \kern -0.75em \lower 0.7ex \hbox{$\sim$}}

\newcommand{\GVec}[1]{\mbox{\boldmath$#1$}}

\def\Vec#1{{\bf #1}}
\def\GVec#1{\mbox{\boldmath $#1$}}

\def\vare{\varepsilon}

\begin{document}

\title{Parity and valley degeneracy in
multilayer graphene}
\author{Mikito Koshino$^{1}$ and Edward McCann$^{2}$}
\affiliation{
$^{1}$Department of Physics, Tokyo Institute of
Technology, 2-12-1 Ookayama, Meguro-ku, Tokyo 152-8551, Japan\\
$^{2}$Department of Physics, Lancaster University, Lancaster, LA1
4YB, UK}

\begin{abstract}
We study spatial symmetry in general $ABA$-stacked multilayer
graphene to illustrate how electronic spectra at the two valleys
are related in a magnetic field.
We show that the lattice of multilayers with an even number of layers,
as well as that of monolayer graphene, satisfy
spatial inversion symmetry, which rigorously guarantees valley degeneracy
in the absence of time-reversal symmetry.
A multilayer with an odd number of layers (three or more) lacks
inversion symmetry, but there is another transformation imposing an
approximate valley degeneracy, which arises because the low-energy Hamiltonian
consists of separate monolayerlike and bilayerlike parts.
We show that an external electrostatic potential generally breaks
valley degeneracy in a magnetic field,
in a markedly different manner in odd and even multilayers.
\end{abstract}


\maketitle

\section{introduction}

\begin{figure}[t]
\centerline{\epsfxsize=0.9\hsize \epsffile{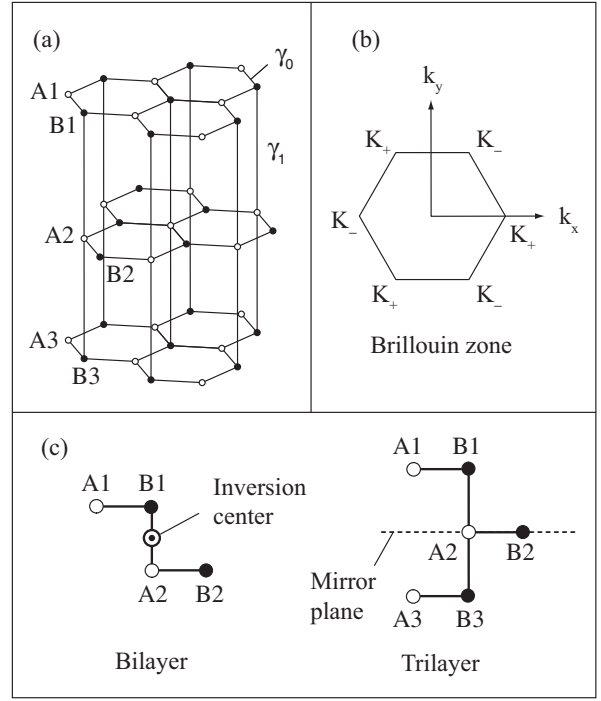}}
\caption{
(a)  Atomic structure of ABA-multilayer graphene.
(b) A schematic of the Brillouin
zone with two inequivalent valleys $K_{\pm}$
(c) Schematic of the lattice of bilayer graphene (left)
and of Bernal-stacked trilayer graphene (right),
with the inversion center for bilayer
and the mirror plane for trilayer, respectively.
}
\label{fig:1}
\end{figure}

The fabrication of individual graphene flakes \cite{novo04}, followed by
an observation of the integer quantum Hall effect
in them \cite{novo05,zhang05,novo06},
triggered an explosion of interest in the electronic properties of graphene.
It was fuelled, in part, by the realization that the low-energy band
structure of a graphene monolayer
consists of two Dirac cones centered at inequivalent corners of
the Brillouin zone, $K_+$ and $K_-$,
which are called valleys [Fig. \ref{fig:1} (b)].
They support chiral quasiparticles
with opposite chirality in each valley, 
and a linear dispersion reminiscent
of the quantum electrodynamics of massless fermions
\cite{semenoff,ando98,gusynin}.
In the presence of time reversal symmetry, the energy spectrum is
degenerate between the different valleys since the
time reversal operation connects electronic states at $K_+$ to those at
$K_-$.
In graphene, not only time reversal symmetry,
but parity, i.e.,
spatial inversion symmetry with respect to the center of a hexagon
\cite{gusynin,zhang06,manes07},
is also able to transform electronic states between valleys.
In the presence of a magnetic field, parity ensures degeneracy
of the electronic spectra at different valleys.

In this paper, we study spatial symmetry in general $ABA$-stacked (Bernal)
multilayer graphene composed of $N$ layers,
to illustrate how the electronic spectra at the two valleys are related
in a magnetic field. In multilayers with even $N$,
including bilayers \cite{novo06,mcc06a,ohta06},
the lattice obeys spatial inversion symmetry
$(x,y,z) \rightarrow (- x, -y , -z)$ similarly to monolayers,
which swaps electronic states between valleys ensuring valley degeneracy
in the absence of time-reversal symmetry.
The picture is different in multilayers with odd $N$, starting
from trilayer graphene ($N=3$), because their lattices do not
satisfy spatial inversion symmetry \cite{latil06,manes07}, Fig.~1(c).
Here we consider an additional
transformation which imposes an approximate valley degeneracy on the
electronic spectra of odd-$N$ multilayers. It arises because the
electronic Hamiltonian may be decomposed into separate monolayer-
and bilayerlike parts \cite{guinea06,part06,kosh_mlg,min08},
with each part satisfying an approximate inversion symmetry corresponding
to that of the real lattice of monolayer or bilayer graphene, respectively.
We predict a peculiar Landau level spectra in bilayer and trilayer
graphenes in the presence of interlayer potential asymmetry,
with unusual structures of broken valley degeneracy
that are markedly different from each other.

\section{Symmetry analysis of the electronic structure}
\label{sec_sym}

We present here general symmetry arguments
without referring to model-specific details, in order
to provide information about the valley degeneracy of Landau levels.
We consider Bernal-stacked multilayer graphene with $N$-layers
in an external uniform magnetic field $\Vec{B}$
with arbitrary direction, and an external potential $U(\Vec{r})$
(other than the lattice potential) where $\Vec{r}=(x,y,z)$.
Without resorting to any approximations,
we can formally express the total Hamiltonian of the system
as $H \left[\mathbf{B} , U(\Vec{r}) \right]$, and we
take $x$ and $y$-axes to be parallel to the graphene layers,
and the $z$ axis in the perpendicular direction.

{\em Even-N multilayers: }
In multilayers with an even number of layers,
the lattice is symmetric with respect to spatial inversion
symmetry $P$ [$(x , y , z) \rightarrow (-x , -y , -z)$]
as shown in Fig. \ref{fig:1} (c),
because the point group of the lattice $D_{3d}$
\cite{latil06,manes07}
($\{ E, 2C_3, 3C_2^{\prime}, i, 2S_6, 3\sigma_d\}$)
can be regarded as a direct product of group
$D_3$ ($\{ E, 2C_3, 3C_2^{\prime}\}$)
with the inversion group $C_i$ ($\{ E, i \}$).
In the absence of $\Vec{B}$, there is a
symmetry with respect to time reversal $T$ ($t \rightarrow -t$).
The influence of $P$ and $T$ on the Hamiltonians of even-$N$ multilayers
is summarized as
\begin{eqnarray}
P:\,\, P H \left[\mathbf{B} , U(\Vec{r}) \right] P^{-1}
&=& H \left[\mathbf{B} , U(-\Vec{r}) \right] \, ,
\label{eq_even_p}\\
T:\,\,\,\,\qquad H^{\ast} \left[\mathbf{B} , U(\Vec{r}) \right]
&=& H \left[-\mathbf{B} , U(\Vec{r}) \right] \, .
\label{eq_even_t}
\end{eqnarray}
The magnetic field $\Vec{B}$ does not change sign under the operation
$P$ because it is an axial vector.
Manes {\em et al} \cite{manes07} used the combined $PT$ operation to show that the Fermi points
are stable with respect to the opening of a gap.

In multilayer graphenes, electronic properties are well described
by a $\Vec{k}\cdot\Vec{p}$ approximation in the vicinity of
$K_+$ and $K_-$ points \cite{kpoints}.
We can consider the $\Vec{k}\cdot\Vec{p}$ Hamiltonian
(and thus eigenstates) at each valley separately,
as long as the potential $U(\Vec{r})$ is smooth in the $xy$-plane
compared to the atomic scale, and the magnetic field is not
too strong as to mix states around the two valleys.
When $\Vec{B}=0$, the time-reversal symmetry in Eq. (\ref{eq_even_t})
ensures valley degeneracy of the electronic spectrum,
because the operation $T$ swaps eigenstates at $K_+$ and those
of $K_-$ through flipping the Bloch factor
as $(e^{i\Vec{K}_{\pm}\cdot \Vec{r}})^* = e^{i\Vec{K}_{\mp}\cdot \Vec{r}}$.
Here $\Vec{K}_\xi(\xi=\pm)$ are the wave numbers corresponding to
$K_\xi$ points \cite{kpoints}.
The operation of spatial inversion $P$ also exchanges $K_+$ and $K_-$
because the point $-\Vec{K}_+$ is equivalent to $\Vec{K}_-$
in the Brillouin zone.
Then the symmetry of Eq. (\ref{eq_even_p})
suggests that the eigenstates of $K_+$ at $[\mathbf{B} , U(\Vec{r})]$
are related to those of $K_-$ at $[\mathbf{B} , U(-\Vec{r})]$.
From this we immediately conclude that
the Landau levels in even-$N$ multilayers are
degenerate in valleys as long as
$U(\Vec{r})=U(-\Vec{r})$, or
the external potential has inversion symmetry
with the same symmetry point as that of the lattice potential.

{\em Odd-N multilayers: }
In odd-$N$ multilayers with $N\geq 3$, the point group
$D_{3h}$ ($\{ E, 2C_3, 3C_2^{\prime}, \sigma_h, 2S_3, 3\sigma_v\}$)
\cite{latil06,manes07} can be regarded as a direct product of the group
$D_3$ ($\{ E, 2C_3, 3C_2^{\prime}\}$) with the reflection group $C_s$ ($\{ E, \sigma_h \}$).
With respect to the even-$N$ multilayers, spatial inversion
is replaced by mirror reflection $\sigma_h$
[$(x , y , z) \rightarrow (x , y , -z)$]
as shown in Fig. \ref{fig:1} (c),
which does not reverse the in-plane electronic momentum.
The influence of $T$ and $\sigma_h$ on the Hamiltonians of odd-$N$ multilayers is summarized as
\begin{eqnarray}
\sigma_h:\,\, \sigma_h H \left[\mathbf{B} , U(x,y,z) \right] \sigma_h^{-1}
&=& H \left[\mathbf{B} , U(x,y,-z) \right] ,
\label{eq_odd_p}\\
T:\,\,\,\,\qquad \qquad H^{\ast} \left[\mathbf{B} , U(\Vec{r}) \right]
&=& H \left[-\mathbf{B} , U(\Vec{r}) \right]  .
\label{eq_odd_t}
\end{eqnarray}
Time reversal symmetry, Eq. (\ref{eq_odd_t}), again
ensures valley degeneracy of the electronic spectrum in the absence of an
external magnetic field.
The mirror reflection symmetry would seem to play the role of
parity. However, it does not transform between states at the two
valleys, and is, therefore, unable to guarantee valley
degeneracy in a magnetic field.
It merely ensures that the spectrum at {\it each} valley
is identical when the potential $U(\Vec{r})$ is inverted with respect to $z=0$.
Actually, we can show that Landau levels in odd-layered multilayer graphenes
are approximately valley-degenerate when $U(\Vec{r})=0$,
by employing the effective mass Hamiltonian described in the following section.

Although we focus on ABA graphene multilayers in this paper,
we point out that ABC (rhombohedral) graphene multilayers,
which have a different layer stacking \cite{haering58,mcclure69},
have inversion symmetry irrespective of their layer number,
and thus the valley degeneracy of Landau levels
is always guaranteed.

\section{Effective mass Hamiltonian}\label{S:emh}

To investigate the Landau level structure in detail
we adopt the effective-mass description of graphite
in the Slonczewski-Weiss-McClure
parameterization \cite{dressel02}.
We consider AB-stacked $N$-layer
multilayer graphene in an external uniform magnetic field
$\Vec{B}$ perpendicular to the layer,
and with an external electrostatic potential $U_j$ at the $j$-th layer,
which is uniform in the in-plane direction.
In a basis with atomic components $\psi_{A1}$, $\psi_{B1}$,
$\psi_{A2}$, $\psi_{B2}$, $\psi_{A3}$, $\psi_{B3}$, $\cdots$, the
multilayer Hamiltonian at the $K_\xi$ valley
\cite{guinea06,part06,lu06,kosh_mlg}
is
\begin{eqnarray}
 {\cal H}^{(\xi)} =
\begin{pmatrix}
 H_1 & V & & & \\
 V^{\dagger} & H_2 & V^{\dagger}& &  \\
  & V & H_3 & V &  & \\
&  & V^{\dagger} & H_4 & V^{\dagger} &  & \\
 & &   & \ddots & \ddots & \ddots &
\end{pmatrix},
\label{eq_H}
\end{eqnarray}
with
\begin{eqnarray}
&& H_i =
\left\{
\begin{array}{ll}
\begin{pmatrix}
 U_i & v \pi^\dagger \\ v \pi & U_i + \delta
\end{pmatrix}
&
(i{\rm : odd})
\\
\begin{pmatrix}
 U_i + \delta & v \pi^\dagger \\ v \pi & U_i
\end{pmatrix}
&
(i{\rm : even})
\end{array}
\right.
\nonumber
\\
&& V =
\begin{pmatrix}
 -v_4\pi^\dagger & v_3 \pi \\ \gamma_1 & -v_4\pi^\dagger
\end{pmatrix}.
\end{eqnarray}
Here  
$\pi = \xi \pi_x + i \pi_y$,
$\xi = \pm 1$ is the valley index
and $\GVec{\pi} = -i\hbar \nabla + e \Vec{A}$
with the vector potential $\Vec{A}$, which gives
the external magnetic field as $\Vec{B} = \nabla\times \Vec{A}$.
Parameter $U_i$ is the potential at the $i$-th layer, which is constant
within each layer, and $\delta$ represents the energy difference
between sites which have neighboring atoms right above or below
them and those sites which do not, and thus it only exists for $N \geq 2$.
Parameter $v$ is the band velocity of monolayer graphene,
which is written as $v = \sqrt{3} a \gamma_0/2\hbar$.
Other velocities are defined as $v_3 = \sqrt{3} a \gamma_3/2\hbar$
and $v_4 = \sqrt{3} a \gamma_4/2\hbar$.
We neglect parameters $\gamma_2$ and $\gamma_5$,
which describe hopping between next-nearest neighboring layers.
They actually break exact valley degeneracy as mentioned later.

For even-$N$ multilayers with $U_j=0$, the Hamiltonian Eq.~(\ref{eq_H})
satisfies
\begin{equation}
\sigma^\dagger_{2N} \mathcal{H}^{(\xi)} \sigma_{2N} = \mathcal{H}^{(-\xi)}
\label{eq_inv_even}
\end{equation}
with $\sigma_{2N}$ being a $2N\times 2N$ matrix,
\begin{equation}
 \sigma_{2N} =
\begin{pmatrix}
&&& 1 \\
&&1 \\
& \rotatebox{90}{$\ddots$} \\
1
\end{pmatrix},
\end{equation}
which is nothing but the inversion symmetry
discussed in the previous section.
In odd-$N$ multilayers with $U_j = 0$,
the Landau levels are degenerate in valleys
due to a different symmetry.
We can show this by employing a unitary transformation
which decomposes the Hamiltonian
of $N$-layer graphene, Eq. (\ref{eq_H}),
into subsystems equivalent
to monolayer and bilayer graphenes.
We construct the basis as \cite{kosh_mlg,kosh09_ssc}
\begin{eqnarray}
 |\phi_{m}^{\rm (X, odd)}\rangle \!\! &=& \!\!
\sum_{j=1}^N f_m(j) |\psi_{X_j}\rangle \nonumber\\
 |\phi_{m}^{\rm (X, even)}\rangle \!\! &=& \!\!
\sum_{j=1}^N g_m(j) |\psi_{X_j}\rangle
\label{eq_basis}
\end{eqnarray}
where $X=A,B$, and
\begin{eqnarray}
 f_m(j) =  \frac{2}{\sqrt{N+1}}
\sin \left(\frac{\pi}{2} j\right)
\cos \left[\frac{m\pi}{2(N+1)} j \right],\\
 g_m(j) =  -\frac{2}{\sqrt{N+1}}
\cos \left(\frac{\pi}{2} j\right)
\sin \left[\frac{m\pi}{2(N+1)} j \right],
\end{eqnarray}
where $j=1,2,\cdots,N$ is the layer index.
The label $m$ is the subsystem index which ranges as
\begin{eqnarray}
 m =
\left\{
\begin{array}{l}
1,3,5, \cdots , N-1, \quad N = {\rm even} \\
0,2,4, \cdots , N-1, \quad N = {\rm odd}
\end{array}
\right.
\label{eq_m}
\end{eqnarray}
The superscript such as (A, odd) indicates that
the wavefunction has an amplitude only on $|A_j\rangle$
with odd $j$'s.

When the Hamiltonian (\ref{eq_H}) with $U_j=0$ is written
in the basis Eq.~(\ref{eq_basis}),
the matrix is block-diagonalized in each $m$.
The case of $m=0$ is special in that
$g_m(j)$ is identically zero, so that
only two bases
$\{|\phi_0^{\rm (A, odd)}\rangle, \, |\phi_{0}^{\rm (B, odd)}\rangle\}$
survive in Eq.~(\ref{eq_basis}).
The submatrix is written for this two component basis as
\begin{eqnarray}
 {\cal H}^{(\xi)}_{m=0} =
\begin{pmatrix}
0 & v\pi^\dagger \\
v\pi &  \delta
\end{pmatrix},
\label{eq_H0}
\end{eqnarray}
which is equivalent to the Hamiltonian of monolayer graphene
except for the diagonal terms containing $\delta$.
For $m \neq 0$, the submatrix for
$\{|\phi_m^{\rm (A, odd)}\rangle, \, |\phi_{m}^{\rm (B, odd)}\rangle,
|\phi_m^{\rm (A, even)}\rangle, \, |\phi_{m}^{\rm (B, even)}\rangle\}$
becomes
\begin{eqnarray}
 {\cal H}^{(\xi)}_{m\neq 0} =
\begin{pmatrix}
0 & v\pi^\dagger &  -\lambda v_4\pi^\dagger & \lambda v_3\pi \\
v\pi & \delta
& \lambda \gamma_1 &  -\lambda v_4\pi^\dagger \\
 -\lambda v_4\pi &  \lambda \gamma_1 & \delta
& v\pi^\dagger \\
 \lambda v_3\pi^\dagger &  -\lambda v_4\pi & v\pi & 0
\end{pmatrix},
\label{eq_Hm}
\end{eqnarray}
where  $\lambda \equiv \lambda_{m}$ is defined by
\begin{eqnarray}
\lambda_m = 2 \cos \kappa_m, \quad
\kappa_m = \frac{\pi}{2}-\frac{m\pi}{2(N+1)}.
\label{eq_lambda}
\end{eqnarray}
Eq.~(\ref{eq_Hm}) is identical to the Hamiltonian of bilayer graphene,
except that interlayer-coupling parameters
$\gamma_1$, $\gamma_3$ and $\gamma_4$ are multiplied by the factor $\lambda$.

Since the decomposed Hamiltonian matrices are analogous to those of
monolayer or bilayer graphene, they obey the corresponding inversion symmetry.
The bilayer-type submatrix, Eq. (\ref{eq_Hm}), obeys
\begin{eqnarray}
 \sigma_4^\dagger \, {\cal H}^{(\xi)}_{m} \, \sigma_4 = {\cal H}^{(-\xi)}_{m},
\label{eq_inv_bi}
\end{eqnarray}
which guarantees valley degeneracy of Landau levels.
In the original basis,
this ``effective'' inversion process Eq.~(\ref{eq_inv_bi})
exchanges the wave amplitudes on odd-$j$th layers
and those on even-$j$th ones.
In even-$N$ multilayers,
the operation Eq.~(\ref{eq_inv_bi})
for each eigenstate
becomes equivalent to the original inversion symmetry
Eq. (\ref{eq_inv_even}) except for a phase factor.
For the monolayer-type submatrix Eq. (\ref{eq_H0}) (exists in odd-$N$)
satisfies
\begin{eqnarray}
 \sigma_2^\dagger \, {\cal H}^{(\xi)}_{m=0} \, \sigma_2
\approx {\cal H}^{(-\xi)}_{m=0},
\label{eq_inv_mono}
\end{eqnarray}
but it is only approximate since
$\delta$ breaks this symmetry, unlike in
the bilayer-type symmetry, Eq. (\ref{eq_inv_bi}).
The operation of Eq. (\ref{eq_inv_mono})
exchanges the wave amplitudes within odd-$j$ layers.

The extra parameters $\gamma_2$ and $\gamma_5$ neglected here
generally mix the states between different $m$'s,
and also appear in diagonal elements within each $m$ \cite{kosh09_ssc}.
In odd-$N$ multilayers, they lift the valley degeneracy
by breaking the effective inversion symmetry  Eq.~(\ref{eq_inv_bi})
through additional matrix elements.
It should be noted that, in even-$N$ multilayers,
valley degeneracy is never influenced
by any extra lattice parameter, because it is protected by
inversion symmetry inherent in the lattice.

\section{Landau level spectra of bilayer and trilayer graphene}

Here, we illustrate the valley degeneracy of the electronic spectra in
even- and odd-$N$ multilayers through an analytical description
of the Landau level spectra of bilayer and trilayer graphene.
We use the magnetic length
$\lambda_{B}=\sqrt{\hbar /(eB)}$ and an energy scale
$\Gamma_B = \sqrt{2\hbar v^2 eB} = \sqrt{2} \hbar v / \lambda_B$
related to the inverse of the magnetic length.
The Landau level spectrum, in a magnetic field of magnitude $B$
perpendicular to the graphene sheet, may be found using
the relation
$(\pi,\pi^\dagger) = (\sqrt{2}\hbar/\lambda_B) (a^\dagger,a)$
for $K_+$, and $(\pi,\pi^\dagger) = (\sqrt{2}\hbar/\lambda_B) (a,a^\dagger)$
for $K_-$ \cite{p+r87},
where $a^\dagger$ and $a$ are raising and lowering operators, respectively,
which operate on the Landau-level wave function $\varphi_n$ as
$a \varphi_n = \sqrt{n}\varphi_{n-1}$ and
$a^\dagger \varphi_n = \sqrt{n+1}\varphi_{n+1}$.

\begin{figure}[t]
\centerline{\epsfxsize=0.8\hsize \epsffile{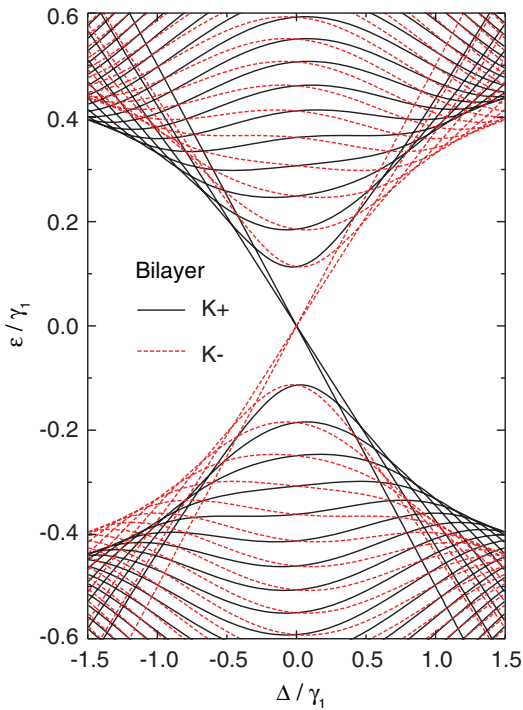}}
\caption{The low-energy Landau level spectrum of bilayer graphene
with $\Gamma_B/\gamma_1=0.3$ at each
valley, plotted as a function of interlayer asymmetry $\Delta$.
}
\label{fig:2}
\end{figure}

{\em Bilayer graphene:} the Hamiltonian of bilayer graphene
\cite{mcc06a,guinea06,mcc06b,kosh_bilayer}
is given by the multilayer Hamiltonian, Eq.~(\ref{eq_H}), with $N=2$.
Here, we consider a simple form of the Hamiltonian that only contains
non-zero parameters $\gamma_0$, describing nearest-neighbor intralayer
hopping, $\gamma_1$, describing the dominant interlayer coupling,
and the layer potential defined as $(U_1,U_2)=\Delta(1,-1)$.
In the absence of interlayer asymmetry, $\Delta = 0$,
the spectrum consists of fourfold (valley and spin) degenerate Landau levels \cite{mcc06a}, except for the
level at zero energy which is eightfold degenerate. The valley degeneracy
is guaranteed by spatial inversion symmetry Eq.~(\ref{eq_even_p}).
Finite $\Delta$ breaks spatial inversion symmetry and it
splits the valley degeneracy of the levels
\cite{mcc06a,castro07,kosh_bilayer_deloc,muchajphys,muchassc,nak09},
as indicated in the numerically-calculated spectrum with a
magnetic field of $\Gamma_B/\gamma_1 = 0.3$,
plotted in Fig.~\ref{fig:2} \cite{chiralnote}.
There we can see that the energy levels of two valleys
are related as $\vare_{K_+}(\Delta) = \vare_{K_-}(-\Delta)$,
due to the inversion symmetry arguments presented in Sec.~\ref{sec_sym}.
The valley splitting $\vare_{K_+}(\Delta) - \vare_{K_-}(\Delta)$
is thus an odd function in $\Delta$,
and generally begins with a term linear in $\Delta$.

In the limit $\{ \varepsilon, \sqrt{n} \Gamma_B , \Delta \}   \ll \gamma_1$,
the Hamiltonian is approximately
described by an effective Hamiltonian \cite{mcc06a} operating in the space of
two-component wave functions $\psi_{A1}$, $\psi_{B2}$
\begin{eqnarray}
\mathcal{H}^{\rm (eff)}_{AB}
&=& -\frac{v^2}{\gamma_1}\left(
\begin{array}{cc}
0 & \left( {\pi }^{\dag }\right) ^{2} \\
{\pi ^{2}} & 0
\end{array} \right) \nonumber \\
&& \hspace{-7mm} + \Delta
\left(
\begin{array}{cc}
1 - 2v^2 \pi^{\dag}\pi / \gamma_1^2 & 0 \\
0 & - 1 + 2v^2 \pi \pi^{\dag} / \gamma_1^2
\end{array}
\right)\, . \label{heff2}
\end{eqnarray}
The energy levels are obtained by assuming the wave function
$\Psi_{n,K_\xi} = ( c_{1} \varphi_{n+\xi}, c_{2} \varphi_{n-\xi})$,
where $\varphi_{m<0}$ is regarded as 0.
They are given by
\begin{eqnarray}
\varepsilon_{n \geq 1} &=&
\pm \sqrt{\frac{\Gamma_B^4}{\gamma_1^2}n(n+1)
+ \Delta^2 \left[
1 - \frac{2\Gamma_B^2}{\gamma_1^2}\left(n+\frac 1 2 \right)
\right]^2}
\nonumber \\
&& - \xi \frac{\Delta \Gamma_B^2}{\gamma_1^2}  \, , \label{bi1} \\
\varepsilon_{0} &=& \xi \Delta \left( 1 - \frac{2 \Gamma_B^2}{\gamma_1^2} \right) \, , \label{bi2} \\
\varepsilon_{-1} &=& \xi \Delta \, . \label{bi3}
\end{eqnarray}
The splitting of the valley degeneracy in $n\geq 1$
(by terms containing $\xi$)
is linear in the asymmetry potential $\Delta$, and also
in the magnetic field $B$.

\begin{figure}[t]
\centerline{\epsfxsize=0.8\hsize \epsffile{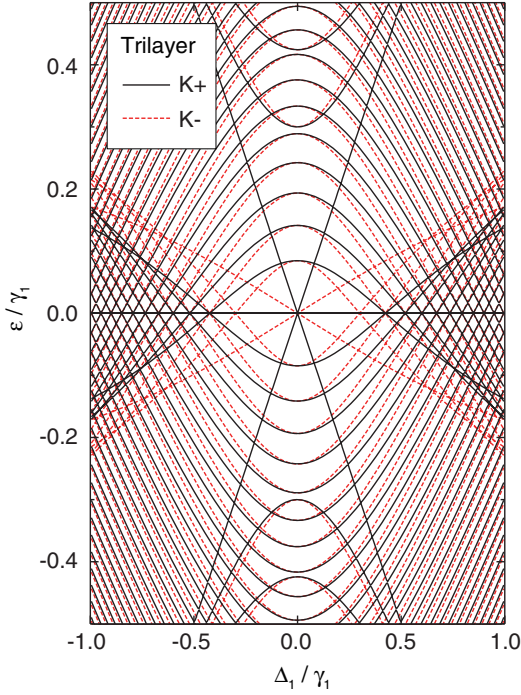}}
\caption{The low-energy Landau level spectrum of ABA-stacked trilayer
 graphene with $\Gamma_B/\gamma_1=0.3$ at each
valley, plotted as a function of interlayer asymmetry $\Delta_1$.
}
\label{fig:3}
\end{figure}

{\em Trilayer graphene:}
The trilayer graphene
\cite{ohta07,guett08,crac08,lu06,guinea06,latil06,part06,kosh_mlg,aoki07,kosh09a}
Hamiltonian is given by the multilayer Hamiltonian, Eq.~(\ref{eq_H}), with $N=3$.
We again retain the parameters $\gamma_0$ and $\gamma_1$,
and set the layer potential as $(U_1,U_2,U_3)=\Delta_1(1,0,-1)$,
to focus our attention on interlayer asymmetry $\Delta_1 = ( U_1 - U_3 )/2$.
As described in Section~\ref{S:emh}, we perform a unitary transformation
to decompose the Hamiltonian into a monolayerlike part, Eq.~(\ref{eq_H0}),
and a bilayerlike part, Eq.~(\ref{eq_Hm}) (such a decomposed trilayer
Hamiltonian is written explicitly in Ref.~\cite{kosh09a}).

In the absence of interlayer asymmetry, $\Delta_1 = 0$,
the spectrum consists of superimposed monolayerlike and bilayerlike spectra \cite{guinea06},
with fourfold (valley and spin) degenerate Landau levels, except for the
level at zero energy which is twelvefold degenerate. The valley degeneracy
is guaranteed by the effective spatial inversion symmetry of the separate,
monolayerlike and bilayerlike, parts of the decomposed Hamiltonian.
Finite $\Delta_1$ breaks the effective spatial inversion symmetry and it
splits the valley degeneracy of the levels
as indicated in the numerically-calculated spectrum at $\Gamma_B/\gamma_1 = 0.3$,
plotted in Fig.~\ref{fig:3} \cite{chiralnote}.
Unlike in bilayer,
the energy spectrum at each valley
is an even function of $\Delta_1$, due to the reflection symmetry
argued in Sec. \ref{sec_sym}.
The valley splitting is therefore quadratic in $\Delta_1$,
except for zero energy where
the different levels are degenerate at each single valley,
and the energy level (and thus splitting)
is allowed to be linear in $\Delta_1$.

In the limit $\{ \varepsilon, \sqrt{n} \Gamma_B , \Delta_1 \}   \ll \gamma_1$,
it is possible to obtain a simplified
description of four electronic bands near zero energy,
by eliminating atomic components associated with bands split away from zero
by energy $\pm \sqrt{2} \gamma_1$,
to obtain a four-component effective Hamiltonian in basis $[\psi_{A1} -
\psi_{A3}]/\sqrt{2}$, $[\psi_{B1} - \psi_{B3}]/\sqrt{2}$,
$[\psi_{A1} + \psi_{A3}]/\sqrt{2}$, $\psi_{B2}$,
\begin{eqnarray}
\mathcal{H}^{\rm (eff)}_{ABA}
= \left(
      \begin{array}{cccc}
        0 & v \pi^{\dag} & \Delta_1 & 0 \\
        v \pi & 0 & 0 & - \frac{\Delta_1 v \pi^{\dag}}{\sqrt{2} \gamma_1} \\
        \Delta_1 & 0 & 0 &  - \frac{v^2(\pi^{\dag})^2}{\sqrt{2} \gamma_1} \\
        0 & - \frac{\Delta_1 v \pi}{\sqrt{2} \gamma_1} &  - \frac{v^2 \pi^2}{\sqrt{2} \gamma_1} & 0 \\
      \end{array}
    \right) , \label{h4b}
\end{eqnarray}

The energy levels are obtained by assuming the wave function
$\Psi_{n,K_\xi} = ( c_{1} \varphi_{n+\xi}, c_{2} \varphi_{n},
c_{3} \varphi_{n+\xi}, c_{4} \varphi_{n-\xi} )$.
For levels with index $n \geq 1$ at valley $K_{+}$,
the Hamiltonian Eq.~(\ref{h4b}) yields
four energy levels for each $n$, given by
\begin{eqnarray}
\varepsilon_{n \geq 1,K+}^{(u)} &=& \pm \sqrt{\Delta_1^2 + (n+1) \Gamma_B^2} \, , \\
\varepsilon_{n \geq 1,K+}^{(l)} &=& \pm \frac{\sqrt{n}\,\Gamma_B}{\sqrt{2}\gamma_1}
\frac{\left[ \Delta_1^2 - (n+1) \Gamma_B^2 \right]}{\sqrt{\Delta_1^2 + (n+1) \Gamma_B^2}} \, .
\end{eqnarray}
For index $n=0$, there are three energy levels,
\begin{eqnarray}
\varepsilon_{0,K+}^{(u)} &=& \pm \sqrt{\Delta_1^2 + \Gamma_B^2} \, , \\
\varepsilon_{0,K+}^{(l)} &=& 0 \, , \label{zp1}
\end{eqnarray}
and, for $n=-1$, two energy levels,
\begin{eqnarray}
\varepsilon_{-1,K+}^{(u)} = \pm \Delta_1 \, . \label{zp2}
\end{eqnarray}
Here, the superscript $(u)$ means that the level
is related to an ``upper'' energy band which is split away from
zero energy by $\Delta_1$ ({\em i.e.} the Landau level energy
approaches $\pm \Delta_1$ as $\Gamma_B \rightarrow 0$),
whereas superscript $(l)$ means that the level
is related to a ``lower'' energy band near
zero energy ({\em i.e.} the Landau level energy
approaches zero as $\Gamma_B \rightarrow 0$).

The Landau level spectrum at the second valley, $K_{-}$,
differs as compared to that at $K_{+}$.
For levels with index $n \geq 1$,
the Hamiltonian Eq.~(\ref{h4b}) again yields
four energy levels for each $n$ given by
\begin{eqnarray}
\varepsilon_{n \geq 1,K-}^{(u)} &=& \pm \sqrt{\Delta_1^2 + n \Gamma_B^2} \, , \\
\varepsilon_{n \geq 1,K-}^{(l)} &=& \pm \frac{\sqrt{n+1}\,\Gamma_B}{\sqrt{2}\gamma_1}
\frac{\left[ \Delta_1^2 - n \Gamma_B^2 \right]}{\sqrt{\Delta_1^2 + n \Gamma_B^2}} \, .
\end{eqnarray}
For index $n=0$, there are two energy levels at $K_{-}$,
\begin{eqnarray}
\varepsilon_{0,K-}^{(l)} = \pm \frac{\Delta_1\Gamma_B}{\sqrt{2}\gamma_1} \, . \label{zm1}
\end{eqnarray}
and, for $n=-1$, one energy level,
\begin{eqnarray}
\varepsilon_{-1,K-}^{(l)} = 0 \, . \label{zm2}
\end{eqnarray}
In the limit $\Delta_1 = 0$,
the levels are degenerate in pairs of
$\varepsilon_{n,K+}^{(u)} = \varepsilon_{n+1,K-}^{(u)}\,(n\geq 0)$,
and $\varepsilon_{n,K+}^{(l)} = \varepsilon_{n,K-}^{(l)}\,(n\geq 1)$,
which form monolayerlike and bilayerlike spectra, respectively.
In the high-field limit, $|\Delta_1| \ll \Gamma_B$,
the valley splitting of those levels
is proportional to $\Delta_1^2/\gamma_1$
but independent of $B$, unlike bilayer graphene.

The other levels give degenerate zero-energy states at $\Delta_1=0$.
For $K_+(K_-)$, one of the levels
$\varepsilon_{-1,K+}^{(u)}(\varepsilon_{0,K-}^{(l)})$
corresponds to the zero-energy level of the monolayerlike spectrum,
whereas
$\varepsilon_{0,K+}^{(l)}(\varepsilon_{-1,K-}^{(l)})$
and one of $\varepsilon_{-1,K+}^{(u)}(\varepsilon_{0,K-}^{(l)})$ correspond
to two zero-energy levels of the bilayer-like spectrum.
This yields an overall twelvefold degeneracy with spin degeneracy
included (as opposed to the
eightfold degeneracy in bilayers and fourfold degeneracy in monolayers).
In the presence of finite interlayer asymmetry, two of the otherwise-zero
levels at each valley are
hybridized and split away from zero [Eq.(\ref{zp2}) at $K_{+}$ and
Eq.(\ref{zm1}) at $K_{-}$], whereas one level at each valley remains
at zero [Eq.(\ref{zp1}) at $K_{+}$ and Eq.(\ref{zm2}) at $K_{-}$].
Zero energy states are formed primarily by different atomic orbitals
at different valleys, and, as in bilayers
\cite{mcc06a,guinea06,kosh09b},
the orbitals in trilayers are on different layers, so that
those levels exhibit quite different dependences
on the interlayer asymmetry between $K_+$ and $K_-$.

In realistic experimental situations,
interlayer potential asymmetry can be produced by
gate-induced electric fields. There the potentials
$U_i$, which are taken as external parameters in the present work,
should be determined self-consistently
including the screening effect of graphene electrons \cite{mcc06b,kosh09a}.
The self-consistent Landau level structure and quantum Hall effect
as a function of gate voltage would be an important future study \cite{fogler}.

It has been pointed out \cite{fogler} that a number of level
crossings will occur in bilayer graphene at finite asymmetry, Fig.~\ref{fig:2}.
We note that the Landau level spectrum of trilayer graphene, Fig.~\ref{fig:3},
has a very rich pattern of level crossings at a range of different values of
energy and asymmetry. The presence of additional terms in the Hamiltonian
 will tend to produce some anti-crossings
in the spectrum. The precise position and nature of level crossings in the
spectra of multilayer graphenes will be the subject of future investigation.

The authors thank T. Ando and M. Mucha-Kruczy\'{n}ski for discussions.
This project has been funded by EPSRC First Grant EP/E063519/1,
the Royal Society, and the Daiwa Anglo-Japanese Foundation,
and by Grants-in-Aid for Scientific Research from the Ministry of Education,
Culture, Sports, Science and Technology, Japan.


\begin{thebibliography}{99}

\bibitem{novo04} K.~S.~Novoselov, A.~K.~Geim, S.~V.~Morozov, D.~Jiang,
Y.~Zhang, S.~V.~Dubonos, I.~V.~Grigorieva, and A.~A.~Firsov
Science \textbf{306}, 666 (2004).

\bibitem{novo05}
K.~S.~Novoselov, A.~K.~Geim, S.~V.~Morozov, D.~Jiang, M.~I.~Katsnelson,
I.~V.~Grigorieva, S.~V.~Dubonos and A.~A.~Firsov,
Nature \textbf{438}, 197 (2005).

\bibitem{zhang05}
Y.~B.~Zhang, Y.~W.~Tan, H.~L.~Stormer, P.~Kim, Nature \textbf{438}, 201 (2005).

\bibitem{novo06}
K.~S.~Novoselov, E.~McCann, S.~V.~Morozov, V.~I.~Fal'ko, M.~I.~Katsnelson,
U.~Zeitler, D.~Jiang, F.~Schedin, and A.~K.~Geim,
Nature Physics \textbf{2}, 177 (2006).

\bibitem{semenoff} G.~W.~Semenoff, Phys. Rev. Lett. \textbf{53},
2449 (1984).

\bibitem{ando98}
T.~Ando, T.~Nakanishi, and R.~Saito, J. Phys. Soc. Jpn.
\textbf{67}, 2857 (1998).

\bibitem{gusynin} V.~P.~Gusynin, S.~G.~Sharapov, J.~Carbotte,
International Journal of Modern Physics B \textbf{21}, 4611 (2007).

\bibitem{zhang06}
Y.~Zhang, Z.~Jiang, J.~P.~Small, M.~S.~Purewal, Y.-W.~Tan,
M.~Fazlollahi, J.~D.~Chudow, J.~A.~Jaszczak, H.~L.~Stormer, and
P.~Kim, Phys. Rev. Lett. {\bf 96}, 136806 (2006).

\bibitem{manes07} J.~L.~Manes, F.~Guinea, and M.~A.~H.~Vozmediano, Phys. Rev. B
\textbf{75}, 155424 (2007).


\bibitem{mcc06a} E.~McCann and V.~I.~Fal'ko, Phys. Rev. Lett. \textbf{96},
086805 (2006).

\bibitem{ohta06} T.~Ohta, A.~Bostwick, T.~Seyller, K.~Horn,
and E.~Rotenberg, Science \textbf{313}, 951 (2006).

\bibitem{latil06} S.~Latil and L.~Henrard, Phys. Rev. Lett. \textbf{97},
036803 (2006).

\bibitem{guinea06} F.~Guinea, A.~H.~Castro~Neto, and N.~M.~R.~Peres, Phys. Rev. B \textbf{73},
245426 (2006).

\bibitem{part06} B.~Partoens and F.~M.~Peeters, Phys. Rev. B \textbf{74},
075404 (2006); {\em ibid.} \textbf{75}, 193402 (2007).

\bibitem{kosh_mlg}
M.~Koshino and T.~Ando,
Phys. Rev. B {\bf 76}, 085425 (2007);
{\em ibid.} {\bf 77}, 115313 (2008).

\bibitem{min08} H.~Min and A.~H.~MacDonald, Phys. Rev. B \textbf{77}, 155416
(2008).

\bibitem{kpoints} Corners of the hexagonal Brillouin zone are located at
wave vector $\mathbf{K}_{\xi }=\xi ({\textstyle\frac{4}{3}}\pi a^{-1},0)$,
where $\xi =\pm 1$ and $a$ is the lattice constant.

\bibitem{haering58}
R. R. Haering, Can. J. Phys. {\bf 36}, 352 (1958).

\bibitem{mcclure69}
J. W. McClure, Carbon \textbf{7}, 425 (1969).

\bibitem{dressel02} M.~S.~Dresselhaus and G.~Dresselhaus,
Adv. Phys. \textbf{51}, 1 (2002).

\bibitem{lu06} C.~L.~Lu, C.~P.~Chang, Y.~C.~Huang, R.~B.~Chen,
and M.~L.~Lin, Phys. Rev. B \textbf{73}, 144427 (2006).

\bibitem{kosh09_ssc}
M.~Koshino and T.~Ando, Solid State Commun. {\bf 149}, 1123 (2009).

\bibitem{p+r87} \textit{\textquotedblleft The Quantum Hall
Effect\textquotedblright }, edited by R.~E.~Prange and S.~M.~Girvin
(Springer-Verlag, New York, 1986).

\bibitem{mcc06b} E.~McCann, Phys. Rev. B \textbf{74}, 161403(R)
(2006).

\bibitem{kosh_bilayer}
M.~Koshino and T.~Ando,
Phys. Rev. B {\bf 73}, 245403 (2006).

\bibitem{castro07} E.V.~Castro {\em et al}, Phys. Rev. Lett. \textbf{99}, 216802
(2007).

\bibitem{kosh_bilayer_deloc}
M. Koshino,
Phys. Rev. B {\bf 78}, 155411 (2008)

\bibitem{muchajphys} M.~Mucha-Kruczy\'{n}ski, D.~S.~L.~Abergel, E.~McCann, and V.~I.~Fal'ko,
J. Phys.:Condens. Matt. \textbf{21}, 344206 (2009).

\bibitem{muchassc} M.~Mucha-Kruczy\'{n}ski, E.~McCann, and V.~I.~Fal'ko,
Solid State Comm. \textbf{149}, 1111 (2009).

\bibitem{nak09} M.~Nakamura, E.~V.~Castro, and B.~D\'ora, arXiv:0910.3469

\bibitem{chiralnote} Note that charge conjugation symmetry \cite{gusynin} of the low-energy effective
Hamiltonians of monolayer, bilayer, and multilayer graphene ensures
electron-hole symmetry of the spectra that, depending on whether $N$ is odd or even, occurs between spectra at
either the same or at opposite valleys, even in the presence of finite
interlayer asymmetry.

\bibitem{kosh09a} M.~Koshino and E.~McCann, Phys. Rev. B
\textbf{79}, 125443 (2009).

\bibitem{ohta07} T.~Ohta, A.~Bostwick, J.~L.~McChesney, T.~Seyller,
K.~Horn, and E.~Rotenberg, Phys. Rev. Lett. \textbf{98}, 206802 (2007).

\bibitem{guett08} J.~G\"uttinger, C.~Stampfer, F.~Molitor, D.~Graf, T.~Ihn, and K.~Ensslin
New Journal of Physics \textbf{10}, 125029 (2008).

\bibitem{crac08} M.~F.~Craciun, S.~Russo, M.~Yamamoto, J.~B.~Oostinga,
A.~F.~Morpurgo, and S.~Tarucha, Nature Nanotechnology \textbf{4}, 383 (2009).

\bibitem{aoki07} M.~Aoki and H.~Amawashi, Solid State Commun.
\textbf{142}, 123 (2007).

\bibitem{kosh09b} M.~Koshino and E.~McCann, Phys. Rev. B
\textbf{80}, 165409 (2009).

\bibitem{fogler} L. M. Zhang, M. M. Fogler, D. P. Arovas, and F. Guinea,
APS March Meeting, 2010 (unpublished),
http://meetings.aps.org/Meeting/MAR10/Event/120799

\end{thebibliography}
\end{document}